\documentstyle[12pt,amssymb]{article}

\def\fnote#1#2{\begingroup\def\thefootnote{#1}\footnote{#2}\addtocounter
{footnote}{-1}\endgroup}

\def\inbar{\vrule height1.5ex width.4pt depth0pt}
\def\IB{\relax{\rm I\kern-.18em B}}
\def\IC{\relax\,\hbox{$\inbar\kern-.3em{\rm C}$}}
\def\ID{\relax{\rm I\kern-.18em D}}
\def\IE{\relax{\rm I\kern-.18em E}}
\def\IF{\relax{\rm I\kern-.18em F}}
\def\IG{\relax\,\hbox{$\inbar\kern-.3em{\rm G}$}}
\def\IH{\relax{\rm I\kern-.18em H}}
\def\II{\relax{\rm I\kern-.18em I}}
\def\IK{\relax{\rm I\kern-.18em K}}
\def\IL{\relax{\rm I\kern-.18em L}}
\def\IM{\relax{\rm I\kern-.18em M}}
\def\IN{\relax{\rm I\kern-.18em N}}
\def\IO{\relax\,\hbox{$\inbar\kern-.3em{\rm O}$}}
\def\IP{\relax{\rm I\kern-.18em P}}
\def\IQ{\relax\,\hbox{$\inbar\kern-.3em{\rm Q}$}}
\def\IR{\relax{\rm I\kern-.18em R}}
\def\IT{\relax{\rm I\kern-.18em T}}
\def\ZZ{\relax{\sf Z\kern-.4em Z}}

\def\a{\alpha}   \def\b{\beta}    \def\g{\gamma}  
\def\e{\epsilon} \def\G{\Gamma}     \def\l{\lambda}
  \def\om{\omega}  \def\Om{\Omega} \def\si{\sigma}

\def\cA{{\cal A}} 
\def\cC{{\cal C}}   
\def\cG{{\cal G}}  \def\cI{{\cal I}} 
  \def\cM{{\cal M}} 
\def\cO{{\cal O}}

   \def\bj{{\bar j}} 
  
  \def\bz{{\bar z}}

\def\bC{{\bar C}}   
 \def\bK{{\bar K}}

 \def\bcG{{\bar \cG}}

\def\bOm{{\bar \Omega}}  
\def\bsi{\bar \sigma} \def\btau{{\bar \tau}}

\def\bmathQ{{\bar \mathQ}}

\def\bfA{{\bf A}} \def\bfE{{\bf E}} \def\bfF{{\bf F}} \def\bfG{{\bf G}}
 \def\bfT{{\bf T}}

\def\hatF{{\hat F}}

\def\rmab{{\rm ab}}

\def\rmVer{{\rm Ver}}

\def\mathC{{\mathbb C}}  \def\mathN{{\mathbb N}}
\def\mathQ{{\mathbb Q}}  \def\mathZ{{\mathbb Z}}

\def\fnote#1#2{\begingroup\def\thefootnote{#1}\footnote{#2}\addtocounter
{footnote}{-1}\endgroup}
\def\beq{\begin{equation}}
\def\eeq{\end{equation}}
\def\bea{\begin{eqnarray}}
\def\eea{\end{eqnarray}}
\def\llea#1{\label{#1}\eea}
\def\lleq#1{\label{#1}\eeq}
\let\nn=\nonumber
\def\tabroom{\hbox to0pt{\phantom{\Huge A}\hss}}
\def\notin{\ \hbox{{$\in$}\kern-.51em\hbox{/}}}

\def\ra{{\rightarrow}}
\def\lra{\longrightarrow}
\def\del{\partial} \def\vphi{\varphi}

  \def\E1Fq{E_1/\IF_q}

\def\rmB{{\rm B}} \def\rmE{{\rm E}} \def\rmF{{\rm F}}
 \def\rmH{{\rm H}}  \def\rmI{{\rm I}}
\def\rmK{{\rm K}}   \def\rmM{{\rm M}}
\def\rmT{{\rm T}}

\def\rmbfT{{\rm \bfT}}

\def\rman{{\rm an}}    \def\rmdim{{\rm dim}}
     \def\rmhom{{\rm hom}}
 \def\rmtor{{\rm tor}}

\def\rmdR{{\rm dR}}     \def\rmCH{{\rm CH}} 
\def\rmRe{{\rm Re}}

\def\rmAut{{\rm Aut}}
\def\rmEnd{{\rm End}} 
\def\rmGal{{\rm Gal}}  \def\rmIso{{\rm Iso}}
\def\rmmod{{\rm mod}}
\def\rmgcd{{\rm gcd}}

\def\Xten{{X^{10}}}
\def\hodgeten{{*_{10}}}  \def\hodgefour{{*_4}}  \def\hodgesix{{*_6}}

\begin{document}

\baselineskip=18pt
\parskip=.01truein
\parindent=0pt

\hfill \phantom{ {\bf PRELIMINARY DRAFT}}


\vskip 1truein

\centerline{\large {\bf Complex Multiplication Symmetry of Black
Hole Attractors}}

\vskip .5truein

\centerline{\sc Monika Lynker\fnote{\star}{Email:
mlynker@iusb.edu}$^1$, Vipul Periwal\fnote{\diamond}{Email:
vipul@gnsbiotech.com}$^2$ and Rolf
Schimmrigk\fnote{\dagger}{Email: netahu@yahoo.com,
rschimmr@kennesaw.edu}$^3$ }

\vskip .4truein

\centerline{{\it $^1$ Indiana University South Bend, South Bend,
IN 46634}}

\vskip .1truein

 \centerline{{\it $^2$ Gene Network Sciences, Ithaca, NY 14850}}

\vskip .1truein

\centerline{{\it $^3$ Kennesaw State University, Kennesaw, GA
30144}}

\vskip 1.1truein

\baselineskip=18pt

\centerline{\large {\bf Abstract}}

We show how Moore's observation, in the context of toroidal
compactifications in type IIB string theory, concerning the
complex multiplication structure of black hole attractor
varieties, can be generalized to Calabi-Yau compactifications with
finite fundamental groups. This generalization leads to an
alternative general framework in terms of motives associated to a
Calabi-Yau variety in which it is possible to address the
arithmetic nature of the attractor varieties in a universal way
via Deligne's period conjecture.

\renewcommand\thepage{}
\newpage

\baselineskip=22.2pt
\parskip=.1truein
\parindent=0pt
\pagenumbering{arabic}

\section{Introduction}

During the past few years number theoretic considerations have
become useful in string theory in addressing a variety of problems
in string theory, such as the understanding of the underlying
conformal field theory of Calabi-Yau manifolds
\cite{s95,s01,su02,gv02}, the nature of black hole attractor
varieties \cite{m98,mm99}, and the behavior of periods under
reduction to finite fields \cite{c00}. Our aim in the present
paper is to further develop and generalize some of the
observations made by Moore in his analysis of the arithmetic
nature of the so-called black hole attractor varieties. The
attractor mechanism \cite{fks95, s96, fk96, fgk97} describes the
radial evolution of vector multiplet scalars of spherical dyonic
black hole solutions in $N=2$ supergravity coupled to abelian
vector multiplets. Under particular regularity conditions the
vector scalars flow to a fixed point in their target space. This
fixed point is determined by the charge of a black hole, described
by a vector $\om$ in the lattice $\Lambda$
 of electric and magnetic charges of the $N=2$ abelian
gauge theory. If the $N=2$ supergravity theory is derived from a
type IIB string theory compactified on a Calabi-Yau space, the
vector multiplet moduli space is described by the moduli space
$\cM$ of complex structures of $X$, and the dyonic charge vector
takes values in the lattice $\Lambda = \rmH^3(X,\mathZ)$.

Moore observed in the context of simple toroidal product
varieties, such as the triple product of elliptic curves $\rmE^3$,
or the product K3$\times \rmE$ of a K3 surface and an elliptic
curve, that the attractor condition determines the complex moduli
$\tau$ of the tori to be given by algebraic numbers in a quadratic
imaginary field $\mathQ(\sqrt{D})$, obtained by adjoining to the
rational numbers $\mathQ$ an imaginary number $\sqrt{D}$, where
$D<0$. This is of interest because for particular points in the
moduli space elliptic curves exhibit additional symmetries, they
admit so-called complex multiplication (CM). For compactifications
with toroidal factors Moore's analysis then appears to indicate an
interesting link between the 'attractiveness' of varieties in
string theory and their complex multiplication properties.

Calabi-Yau varieties with elliptic factors are very special
because they have infinite fundamental group, a property not
shared by Calabi-Yau manifolds in general. Other special features
of elliptic curves are not present in general either. In
particular Calabi-Yau spaces are not abelian varieties and they do
not, in any obvious fashion, admit complex multiplication
symmetries. Hence it is not clear how Moore's observations can be
generalized. It is this problem which we wish to address in the
present paper. In order to do so we adopt a cohomological approach
and view the modular parameter of the elliptic curve as part of
the primitive cohomology. In the case of elliptic curves $\rmE$
this is simply a choice of view because there exists an
isomorphism between the curve itself and its Jacobian defined by
$J(\rmE)=\rmH^1(\rmE,\mathC)/\rmH^1(\rmE,\mathZ)$ described by the
Abel-Jacobi map $j: \rmE \ra J(\rmE)$. These varieties are
abelian.

The Jacobian variety of an elliptic (or more general) curve has a
natural generalization to higher dimensional varieties, defined by
the intermediate Jacobian of Griffiths. It would be natural to use
Griffiths' construction in an attempt to generalize the elliptic
results described above. In general, however, the intermediate
Jacobian is not an abelian variety and does not admit complex
multiplication. For this and other reasons we will proceed
differently by constructing a decomposition of the intermediate
cohomology of the Calabi-Yau and using this decomposition to
formulate a generalization of the concept of complex
multiplication of black hole attractor varieties. To achieve this
we formulate complex multiplication in this more general context
by analyzing in some detail the cohomology group $\rmH^3(X)$  of
weighted Fermat hypersurfaces.

The paper is organized as follows. In order to make the
presentation more self-contained we briefly review in Section 2
the physical setting of black hole attractors in type IIB
theories, as well as Moore's solution of the $\rmK3 \times \rmE$
solution of the attractor equations. In Section 3 we describe the
necessary background of abelian varieties, and in Section 4 we
show how abelian varieties can be derived from Calabi-Yau
hypersurfaces by showing that the cohomology of such varieties can
be constructed from the cohomology of curves embedded in these
higher dimensional varieties. This leads us to abelian varieties
defined by the Jacobians of curves. Such abelian varieties do not,
in general, admit complex multiplication. What can be shown,
however, is that Jacobians of projective Fermat curves split into
abelian factors which {\it do} admit complex multiplication. We
briefly describe this construction and generalize the discussion
to curves of Brieskorn-Pham type. Combining these results shows
that we can define the complex multiplication type of Calabi-Yau
varieties with finite fundamental groups via the CM types of their
underlying Jacobians. In Section 5 we indicate some of the
arithmetic consequences for Calabi-Yau varieties that derive from
the emergence of abelian varieties with complex multiplication in
the context of black hole attractors.

In the process of our analysis we will recover the same fields
which Moore uncovered by considering the fields generated by
periods of higher dimensional varieties. Even though our approach
is very different from Moore's, it is not completely unexpected
that we should be able to recover the field of periods by
considering the complex multiplication type. The reason for this
is a conjecture of Deligne \cite{d79} which states that the field
determined by the periods of a critical motive is determined by
its L-function. Because Deligne's conjecture is important for our
general view of the issue at hand, we briefly describe this
conjecture in  Section 6 in order to provide the appropriate
perspective. Deligne's conjecture is in fact a theorem in the
context of projective Fermat hypersurfaces \cite{b86}, but has not
been proven in the context of weighted hypersurfaces. Our results
in essence can be viewed as support of this conjecture even in
this more general context. In Section 7 we summarize our results
and indicate possible generalizations.

\vskip .4truein

\section{Arithmetic of Elliptic Attractor Varieties}

\subsection{Attractor Varieties}

In this paper we consider type IIB string theory compactified on
Calabi-Yau threefold varieties. The field content of the string
theory in 10D space $\Xten$ splits into two sectors according to
the boundary conditions on the world sheet. The Neveu-Schwarz
fields are given by the metric $g \in \G(\Xten, \rmT^*\Xten\otimes
\rmT^*\Xten)$, an antisymmetric tensor field $B \in \G(\Xten,
\Om^2)$ and the dilaton scalar $\phi \in C^{\infty}(\Xten, \IR)$.
The Ramond sector is spanned by even antisymmetric forms $A^p \in
\G(\Xten, \Om^p)$ of rank $p$ zero, two, and four. Here $\Om^p
\lra X$ denotes the bundle of $p$-forms over the variety $X$.

In the context of the black hole solutions considered in
\cite{fks95} the pertinent sectors are given by the metric and the
five-form field strength $\mathbf{F}$ of the Ramond-Ramond 4-form
$\bfA^4$. The metric is assumed to be a static spherically
symmetric spacetime which is asymptotically Minkowskian and
describes an extremally charged black hole, leading to the ansatz
\beq ds^2 =-e^{2U(r)} dt\otimes dt + e^{-2U(r)}(dr\otimes dr +
r^2\si_2),\eeq where $r$ is the spatial three dimensional radius,
$\si_2$ is the 2D angular element, and the asymptotic behavior is
encoded via $e^{-U(r)} \ra \infty$ for $r \ra \infty$. The
expansion of the five-form $\mathbf{F}$ leads to a number of
different 4D fields, the most important in the present context
being the field strengths  $F^L$ of four dimensional
 abelian fields, the number
of which depends on the dimension of the cohomology group
$\rmH^3(X)$ via \beq \bfA^4_{\mu mnp}(x,y) = \sum_L
A^{4L}_{\mu}(x) \om^L_{mnp}(y),\eeq where
$\{\om_L\}_{L=1,...,b_3}$ is a basis of $\rmH^3(X)$. This is
usually written in terms of a symplectic basis $\{\a_a,
\b^a\}_{a=0,...,h^{2,1}}$, for which $\int_X \a_a \wedge \b^b
=\delta_a^b$, leading to an expansion of the field strength of the
form \beq \bfF (x,y) = \bfF^a(x) \wedge \a_a - \bfG_a(x) \wedge
\b^a. \eeq Being a five-form in ten dimensions the field strength
$\bfF$ admits (anti)self-duality constraints with respect to Hodge
duality, $\bfF = \pm \hodgeten \bfF$. The ten dimensional Hodge
operator $\hodgeten$ factorizes into a 4D and a 6D part $\hodgeten
= \hodgefour \hodgesix$. A solution to the anti-selfduality
constraint in 10D as well as the Bianchi identity $d\bfF =0$ can
be obtained by considering \cite{m98} \beq \bfF = \rmRe \left(\bfE
\wedge (\om^{2,1} + \om^{0,3} ) \right), \eeq where \cite{f97}
\beq \bfE \equiv q \sin \theta d\theta \wedge d\phi - i q
\frac{e^{2U(r)}}{r^2} dt \wedge dr \eeq is a 2-form for which the
four dimensional Hodge duality operator leads to $\hodgefour \bfE
=i \bfE$. The 6D Hodge dual on $\rmH^3(X)$ is defined by \beq
\hodgesix = -i \Pi^{3,0} \oplus i \Pi^{2,1} \oplus -i \Pi^{1,2}
\oplus i \Pi^{0,3}, \eeq leading to a purely imaginary duality
transformation of the internal part of $\bfE$.

Two standard maneuvers to derive the dynamics of a string
background configuration are provided by the reduced IIB effective
action with a sort of small superspace ansatz \cite{fgk97}, and
the supersymmetry variation constraints of the fermions in
nontrivial backgrounds, in particular the gravitino and gaugino
variations. We adopt the notation of \cite{cd90}. Defining an
inner product $<\cdot , \cdot>$ on $\rmH^3(X)$ via \beq <\om,
\eta> = \int_X \om \wedge \eta, \eeq the gravitino equation
involves the integrated version of the 5-form field strength
defined as \cite{bcdffrsv95} \cite{cdf95} \beq \bfT^{-} = e^{K/2}
<\Om, \bfF^{-}> =  e^{K/2} \left( \cG_a \bfF^{a-}(x) -
z^a\bfG_a^{-}(x) \right), \eeq with K\"ahler potential \beq e^{-K}
= i<\Om, \bOm> = -i(z^a \bcG_a - \bz^a \cG_a).\eeq Here the second
equation is written in terms
of the periods \bea z^a &=& <\Om, \b^a> = \int_{A^a}\Om \nn \\
\cG_a &=& <\Om, \a_a> = \int_{B_a} \Om,\eea with respect to a
symplectic homological basis $\{A^a, B_a\}_{a=0,...,h^{2,1}}
\subset \rmH_3(X)$ which is dual to the cohomological basis
$\{\a_a, \b^a\}_{a=0,...,h^{2,1}} \subset \rmH^3(X)$. The
holomorphic three-form thus can be expanded as \beq \Om = z^a\a_a
-\cG_a \b^a.\eeq

The supersymmetry transformation of the gravitino $\psi^A =
\psi_{\mu}^A dx^{\mu}$ can then be written in terms of the
components of $\bfT^-$ as \beq \delta \psi^A = D \varepsilon^A +
dx^{\mu}\rmT^{-}_{\mu \nu} \gamma^{\nu} (\epsilon
\varepsilon)^A,\eeq where $\g^{\mu}$ denotes the covariant Dirac
matrices and \beq D = dx^{\mu} D_{\mu} = dx^{\mu} \left(
\del_{\mu} - \frac{1}{4}\om_{\mu}^{ab}\g_{ab} + i Q_{\mu}\right)
\eeq is a derivative covariant with respect to both the Lorentz
and the K\"ahler transformations in terms of the spin connection
$\om_{\mu}^{ab}$ and the K\"ahler connection $Q_{\mu}$. The
variation of the gaugino of the abelian multiplets takes the form
\beq \delta \l^{iA} = i\g^{\mu} \del_{\mu} z^i\e^A +
\frac{i}{2}G^{i-}_{\mu \nu} \g^{\mu \nu} (\e \varepsilon )^A.\eeq

Plugging these ingredients into the supersymmetry transformation
behavior of the gravitino and the gaugino fields, and demanding
that the vacuum remains fermion free, leads to  the following
equations for the moduli $z^i$ and the spacetime
function $U(r)$ \bea \frac{dU}{d\rho} &=& - e^U|Z| \nn \\
\frac{dz^i}{d\rho} &=& -2e^U g^{i\bj} \del_{\bj} |Z|,\eea where
$\rho = 1/r$, $g_{i\bj} = \del_i \del_{\bj} K$ is the metric
derived from the K\"ahler potential $K$, and \beq Z(\Gamma) =
e^{K/2} \int_{\Gamma} \Om = e^{K/2} \int_X \eta_{\Gamma} \wedge
\Om \eeq is the central charge of the cycle $\Gamma \in \rmH_3(X)$
with Poincare dual $\eta_{\Gamma} \in \rmH^3(X)$. To make the
moduli and charge dependence of the central charge explicit one
can alternatively view $Z(\Gamma)$ as the integral of the
graviphoton form \beq Z(z^a, p^a, q_a) = \int_{S^2} \rmbfT^- =
e^{K/2}\left(\cG_a p^a -z^a q_a\right)\eeq in terms of the charges
\bea p^a &=& \int_{S^2} \bfF^{a-} \nn \\
     q_a &=& \int_{S^2} \bfG^-_a.\eea

The fixed point condition of the attractor equations can be
written in a geometrical way as the Hodge condition \beq
\rmH^3(X,\mathZ) \ni \om = \om^{3,0} + \om^{0,3}. \eeq Writing
$\om^{3,0} =-i\bC \Om$ this can be formulated as
\bea ip^a &=& \bC z^a - C \bz^a \nn \\
     iq_a &=& \bC \cG_a - C\bcG_a,\llea{fxdpnteq}
where $C=e^{K/2}Z$.

The system (\ref{fxdpnteq}) describes a set of $b_3(X)$ charges
$(p^a,q_a)$ determined by the physical 4-dimensional input, which
in turn determines the complex periods of the Calabi-Yau variety.
Hence the system should be solvable. The interesting structure of
the fixed point which emerges is that the central charges are
determined completely in terms of the charges of the
four-dimensional theory. As a consequence the 4D geometry is such
that the horizon is a moduli independent quantity. This is
precisely as expected because the black hole entropy should not
depend on adiabatic changes of the environment \cite{lw95}.

\subsection{Arithmetic of Attractor Elliptic Curves}

In reference \cite{m98} Moore noted that two types of solutions of
the attractor equations have particularly interesting properties.
The first of these is provided by the triple product of a torus,
while the second is a product of a K3 surface and a torus. Both
solutions are special in the sense that they involve elliptic
curves. In the case of the product threefold $X=\rmK3 \times \rmE$
the simplifying feature is that via K\"unneth's theorem one finds
$\rmH^3(\rmK3 \times \rmE) \cong \rmH^2(\rmK3) \otimes
\rmH^1(\rmE)$, and therefore the cohomology group of the threefold
in the middle dimension is isomorphic to two copies of the
cohomology group $\rmH^2(\rmK3)$ because $\rmH^1(\rmE)$ is two
dimensional. The attractor equations for such threefolds have been
considered in \cite{adf96}. The resulting constraints determine
the holomorphic form of both factors in terms of the charges
$(p,q)$ of the fields. The complex structure $\tau$ of the
elliptic curve $\rmE_{\tau} = \mathC/(\mathZ+\tau \mathZ)$ is
solved as \beq \tau_{p,q} = \frac{p\cdot q +\sqrt{D_{p,q}}}{p^2},
\eeq where $D_{p,q}=(p\cdot q)^2 - p^2q^2$ is the discriminant of
a BPS state labelled by \beq \om = (p,q) \in \rmH^3(\rmK3\times
\rmE,\mathZ).\eeq The holomorphic two form on K3 is determined as
$\Om^{2,0} = \cC (q-\btau p)$, where $\cC$ is a constant.

Moore makes the interesting observation that this result is known
to imply that the elliptic curve determined by the attractor
equation is distinguished by exhibiting a particularly symmetric
structure. Elliptic curves are groups and therefore one can
consider the endomorphism algebra $\rmEnd(\rmE)$. This algebra can
take one of three forms. Generally, $\rmEnd(\rmE)$ is just the
ring $\mathZ$ of rational integers. For special curves however
there are two other possibilities for which $\rmEnd(\rmE)$ is
either an order of a quadratic imaginary field $F$, i.e. it is a
subring $\cO_F$ which generates $F$ as a $\mathQ$-vector space and
is finitely generated as a $\mathZ-$module, or it is a maximal
order in a quaternion algebra. The latter possibility can occur
only when the field $K$ over which $\rmE$ is defined has positive
characteristics. Elliptic curves are said to admit complex
multiplication if the endomorphism algebra is strictly larger than
the ring of rational integers\fnote{1}{A brief review of the
arithmetic theory of elliptic curves can be found in \cite{s97}. A
more extended source \cite{s94}.}.

The point here is that the property of complex multiplication
appears if and only if the $j$-invariant $j(\tau)$ of an elliptic
curve $E_{\tau} = \mathC/(\mathZ +\tau\mathZ)$ is an algebraic
integer, i.e. it solves a polynomial equation with rational
coefficients such that the coefficient of the leading term is
unity. This happens if and only if the modulus $\tau$ is an
imaginary quadratic number, i.e. it solves an equation $A\tau^2
+B\tau +C=0$ for $A,B,C \in \mathZ$. The $j$-invariant of the
elliptic curve $\rmE_{\tau}$ can be defined in terms of the
Eisenstein series \beq E_k(\tau) = \frac{1}{2} \sum_{\stackrel{m,n
\in \mathZ}{m,n~{\rm coprime}}} \frac{1}{(m\tau +n)^k} \eeq as
\beq j(\tau) = \frac{E_4(\tau)^3}{\Delta(\tau)},\eeq
 where $1728 \Delta(\tau) = E_4(\tau)^3 - E_6(\tau)^2$.
In general $j(\tau)$ does not take algebraic values, not to
mention values in an imaginary quadratic field. Even if $\tau$ is
an algebraic number will $j(\tau)$ be a transcendental number
unless $\tau$ is imaginary quadratic. Thus we see that in the
framework of toroidal compactification the solutions of the
attractor equations can be characterized as varieties which are
unusually symmetric and which admit complex multiplication by a
quadratic imaginary field $F = \mathQ(i\sqrt{|D|})$.

Once this is recognized several classical results about elliptic
curves with complex multiplication are available to illuminate the
nature of the attractor variety. Exploring these consequences is
of interest because it provides tools that allow a
characterization of attractor varieties that involve elliptic
factors. The nature of attractor varieties without elliptic
factors is at present not understood, and finding generalizations
of the arithmetic results obtained in the elliptic context
provides a framework in which Calabi-Yau varieties with finite
fundamental groups (which may be trivial) can be explored.

One of the important number theoretic results associated to
elliptic curves with complex multiplication is that the extension
$F(j(\tau))$ obtained by adjoining the $j$-value to $F$ is the
maximal unramified extension of $F$ with an abelian Galois group,
i.e. the Hilbert class field. Geometrically there is a Weierstrass
model, i.e. a projective embedding of the elliptic curve of the
form \beq y^2 + a_1xy +a_3y = x^3 + a_2x^2 + a_4x + a_6 \eeq that
is defined over this extension $F(j(\tau))$.

Even more interesting is that it is possible to construct from the
geometry of the elliptic curve the maximal abelian extension
$F_{\rmab}$ of $F$ by considering the torsion points
$\rmE_{\rmtor}$ on the curve $\rmE$, i.e. points of finite order
with respect to the group law. The field $F(j(\tau),
\rmE_{\rmtor})$, defined by the $j-$function and the torsion
points, is in general not an abelian extension of $F$, but
contains the maximal abelian extension $F_{\rmab}$. It is possible
to isolate $F_{\rmab}$ by mapping the torsion points via the Weber
function \beq \Phi_{\rmE}: \rmE \lra \IP_1, \eeq of the curve
$\rmE$, whose definition depends on the automorphism group of the
curve. Assuming that the characteristic of the field $F$ is
different from 2 or 3, the elliptic curve can be embedded via the
simplified Weierstrass form \beq y^2=x^3 +Ax +B. \eeq If the
discriminant \beq \Delta = -16(4A^3+27B^2) \eeq does not vanish
the elliptic curve is smooth, and the automorphism group
$\rmAut(\rmE)$ can be shown to take one of the following forms,
depending on the value of the $j-$invariant \beq \rmAut(\rmE) =
\left\{
\begin{tabular}{l l}
     $\{\pm 1\}$  & if $j(\rmE) \neq 0 ~{\rm or}~ 1728$, i.e. $AB\neq 0$ \tabroom \\
     $\{\pm 1,\pm i\}$  & if $j(\rmE)=1728$, i.e. $B=0$  \tabroom  \\
     $\{\pm 1,\pm \xi_3,\pm \xi_3^2\}$   & if $j(\rmE)=0$, i.e. $A=0$, \tabroom  \\
\end{tabular} \right\}
\eeq where $\xi_3=e^{2\pi i/3}$ is a primitive third root of
unity.

The Weber function can then be defined as \beq \Phi_{\rmE}(p) =
\left\{
\begin{tabular}{l l}
     $\frac{AB}{\Delta}x(p)$
        & if $j(\rmE) \neq 0 ~{\rm or}~ 1728$ \tabroom \\
     $\frac{A^2}{\Delta}x^2(p)$  & if $j(\rmE)=1728$ \tabroom  \\
     $\frac{B}{\Delta}x^3(p)$   & if $j(\rmE)=0$. \tabroom  \\
  \end{tabular} \right\}
\eeq and the Hilbert class field $F(j(\tau))$ can be extended to
the maximal abelian extension $F_{\rmab}$ of $F$ by adjoining the
Weber values of the torsion points \beq F_{\rmab} = F(j(\tau),
\{\Phi_{\rmE}(t)~|~t\in \rmE_{\rmtor}\}).\eeq

We see from these results that the attractor equations pick out
special elliptic curves with an enhanced symmetry group. This is
an infinite discrete group which in turn leads to a rich
arithmetic structure. It is this set of tools which we wish to
generalize to the framework of Calabi-Yau varieties proper, i.e.
those with finite fundamental group.

\vskip .4truein

\section{Abelian Varieties with Complex Multiplication}

We first review some pertinent definitions of abelian varieties.
An abelian variety over some number field $K$ is a smooth,
geometrically connected, projective variety which is also an
algebraic group with a group law $A\times A \lra A$  defined over
$K$. A concrete way to construct abelian varieties is via complex
tori $\mathC^n/\Lambda$ with respect to some lattice $\Lambda$,
that is not necessarily integral, and admits a Riemann form. The
latter is defined as an $\IR$-bilinear form $<,>$ on $\mathC^n$
such that $<x,y>$ takes integral values for all $x,y\in \Lambda$,
and satisfies the relations  $<x,y>=-<y,x>$. Furthermore, $<x,iy>$
is a positive symmetric form, not necessarily non-degenerate. The
result then is that a complex torus $\mathC^n/\Lambda$ has the
structure of an abelian variety if and only if there exists a
non-degenerate Riemann form on $\mathC^n/\Lambda$.

A special class of abelian varieties are those of CM type,
so-called complex multiplication type. The reason these varieties
are special is because, as in the lower dimensional case of
elliptic curves, certain number theoretic question can be
addressed in a systematic fashion for this class. Consider a
number field $F$ over the rational numbers $\mathQ$ and denote by
$[F:\mathQ]$ the degree of the field $F$ over $\mathQ$, i.e. the
dimension of $F$ over the subfield $\mathQ$. An abelian variety
$A$ of dimension $n$ is called a CM$-$variety if there exists an
algebraic number field $F$ of degree $[F:\mathQ]=2n$ over the
rationals $\mathQ$ which can be embedded into the endomorphism
algebra $\rmEnd(A) \otimes \mathQ$ of the variety. More precisely,
a CM variety is a pair $(A,\theta)$ with $\theta: F \lra \rmEnd(A)
\otimes \mathQ$ an embedding of $F$. It follows from this that the
field $F$ necessarily is a CM field, i.e. a totally imaginary
quadratic extension of a totally real field. The important
ingredient here is that restriction to $\theta(F) \subset
\rmEnd(A)\otimes \mathQ$ is equivalent to the direct sum of $n$
isomorphisms $\phi_1,...,\phi_n \in \rmIso(F,\mathC)$ such that
$\rmIso(F,\mathC) = \{\phi_1,...,\phi_n, \rho\phi_1,....,\rho
\phi_n\}$, where $\rho$ denotes complex conjugation. These
considerations lead to the definition  of calling the pair
$(F,\{\phi_i\})$ a CM type, in the present context, the CM type of
a CM variety $(A,\theta)$.

The context in which these concepts will appear in this paper is
provided by varieties which have complex multiplication by a
cyclotomic field $F=\mathQ(\mu_n)$, where $\mu_n$ denotes the
cyclic group generated by a primitive $n$'th root of unity
$\xi_n$. The field $\mathQ(\mu_n)$ is the imaginary quadratic
extension of the totally real field $\mathQ(\xi_n+ {\bar \xi}_n) =
\mathQ(\cos(2\pi/n))$ and therefore is a CM field. The degree of
$\mathQ(\mu_n)$ is given by $[\mathQ(\mu_n):\mathQ]=\vphi(n)$,
where $\vphi(n)=\# \{m\in \mathN ~|~m<n, ~\rmgcd(m,n)=1\}$ is the
Euler function. Hence the abelian varieties we will encounter will
have complex dimension $\vphi(n)/2$. Standard references for
abelian varieties with complex multiplication have been provided
by Shimura \cite{st61, s71, s98}.

In the following parts we first reduce the cohomology of the
Brieskorn-Pham varieties to that generated by curves and then
analyze the structure of the resulting weighted curve Jacobians.

\vskip .4truein

\section{Abelian Varieties from Brieskorn-Pham type hypersurfaces}

\subsection{Curves and the cohomology of threefolds}

The difficulty of higher dimensional varieties is that there is no
immediate way to recover abelian varieties, thus making it
non-obvious how to generalize the concept of complex
multiplication from one-dimensional Calabi-Yau varieties, which
are abelian varieties, to K3 surfaces and higher dimensional
spaces. As a first step we need to disentangle the Jacobian of the
elliptic curve from the curve itself. This would lead us to the
concept of the middle-dimensional cohomology, more precisely the
intermediate (Griffiths) Jacobian which is the appropriate
generalization of the Jacobian of complex curves. The problem with
this intermediate Jacobian is that it is not, in general, an
abelian variety.

We will show now that it is possible nevertheless to recover
abelian varieties as the basic building blocks of the intermediate
cohomology in the case of weighted projective hypersurfaces. The
basic reason for this is that the cohomology $\rmH^3(X)$ for these
varieties decomposes into the monomial part and the part coming
from the resolution. The monomial part of the intermediate
cohomology can be obtained from the cohomology of a projective
hypersurface of the same degree by realizing the weighted
projective space as a quotient variety with respect to a product
of discrete groups determined by the weights of the coordinates.
For projective varieties \beq X_d^n =\left\{(z_0,...,z_{n+1})\in
\IP_{n+1}~|~z_0^d + \cdots + z_{n+1}^d = 0\right\} \subset
\IP_{n+1}\eeq it was shown in \cite{sk79} that the intermediate
cohomology can be determined by lower-dimensional varieties in
combination with Tate twists by reconstructing the higher
dimensional variety $X_d^n$ of degree $d$ and dimension $n$ in
terms of lower dimensional varieties $X_d^r$ and $X_d^s$ of the
same degree with $n=r+s$. Briefly, this works as follows. The
decomposition of $X_d^n$ is given as \beq X_d^{r+s} ~\cong ~
B_{Z_1,Z_2}\left(\left(\pi_Y^{-1}(X_d^r \times
X_d^s)\right)/\mu_d\right), \eeq which involves the following
ingredients.  \hfill \break (1) $\pi_Y^{-1}(X_d^r\times X_d^s)$
denotes the blow-up of $X_d^r\times X_d^s$ along the subvariety
\beq Y = X_d^{r-1}\times X_d^{s-1} \subset X_d^r \times X_d^s.\eeq
The variety $Y$ is determined by the fact that the initial map
which establishes the relation between the three varieties
$X_d^{r+s},X_d^r, X_d^s$ is defined on the ambient spaces as \beq
((x_0,...,x_{r+1}),(y_0,...,y_{s+1}) ~\mapsto~
(x_0y_{s+1},...,x_ry_{s+1},x_{r+1}y_0,...,x_{r+1}y_s).\eeq This
map is not defined on the subvariety $Y$. \hfill \break (2)
$\pi_Y^{-1}(X_d^r \times X_d^s)/\mu_d$ denotes the quotient of the
blow-up $\pi_Y^{-1}(X_d^r \times X_d^s)$ with respect to the
action of
$$\mu_d \ni \xi: ((x_0,...,x_r,x_{r+1}),(y_0,...,y_s,y_{s+1})) ~\mapsto ~
((x_0,...,x_r,\xi x_{r+1}), (y_0,...,y_s,\xi y_{s+1})).$$ (3)
$B_{Z_1,Z_2}\left(\left(\pi_Y^{-1}(X_d^r \times
X_d^s)\right)/\mu_d\right)$ denotes the blow-down in
$\pi_Y^{-1}(X_d^r \times X_d^s)/\mu_d$ of the two subvarieties
$$ Z_1=\IP_r \times X_d^{s-1},~~~~~~Z_2 = X_d^{r-1} \times \IP_s.$$

This construction leads to an iterative decomposition of the
cohomology which takes the following form. Denote the Tate twist
by \beq \rmH^i(X)(j):=\rmH^i(X)\otimes W^{\otimes j}\eeq with
$W=\rmH^2(\IP_1)$ and let $X_d^{r+s}$ be a Fermat variety of
degree $d$ and dimension $r+s$. Then \bea \rmH^{r+s}(X_d^{r+s}) &
\oplus &\sum_{j=1}^r \rmH^{r+s-2j}(X_d^{r-1})(j) \oplus
\sum_{k=1}^s
\rmH^{r+s-2k}(X_d^{s-1})(k) \nn \\
 & & ~~~~ \cong \rmH^{r+s}(X_d^r\times
X_d^s)^{\mu_d} \oplus \rmH^{r+s-2}(X_d^{r-1}\times
X_d^{s-1})(1).\eea This allows us to trace the cohomology of
higher dimensional varieties to that of curves.

Weighted projective hypersurfaces can be viewed as resolved
quotients of hypersurfaces embedded in ordinary projective space.
The resulting cohomology has two components, the invariant part
coming from the projection of the quotient, and the resolution
part. As described in \cite{cls90}, the only singular sets on
arbitrary weighted hypersurface Calabi-Yau threefolds are either
points or curves. The resolution of singular points contributes to
the even cohomology group $\rmH^2(X)$ of the variety, but does not
contribute to the middle-dimensional cohomology group $H^3(X)$.
Hence we need to be concerned only with the resolution of curves
(see e.g. \cite{s87}). This can be described for general CY
hypersurface threefolds as follows. If a discrete symmetry group
$\mathZ/n\mathZ$ of order $n$ acting on the threefold leaves
invariant a curve then the normal bundle has fibres $\mathC_2$ and
the discrete group induces an action on these fibres which can be
described by a matrix \beq \left(\matrix{\a^{mq} &0 \cr 0
&\a^m\cr}\right),\eeq where $\a$ is an $n$'th root of unity and
$(q,n)$ have no common divisor. The quotient
$\mathC_2/(\mathZ/n\mathZ)$ by this action has an isolated
singularity which can be described as the singular set of the
surface in $\mathC_3$ given by the equation \beq
S=\{(z_1,z_2,z_3)\in \mathC_3~|~z_3^n=z_1z_2^{n-q}\}.\eeq The
resolution of such a singularity is completely determined by the
type $(n,q)$ of the action by computing the continued fraction of
$\frac{n}{q}$ \beq \frac{n}{q}= b_1 - \frac{1}{b_2 -
\frac{1}{\ddots - \frac{1}{b_s}}} \equiv [b_1,...,b_s].\eeq The
numbers $b_i$ specify completely the plumbing process that
replaces the singularity and in particular determine the
additional generator to the cohomology $\rmH^*(X)$ because the
number of $\IP_1$s introduced in this process is precisely the
number of steps needed in the evaluation of
$\frac{n}{q}=[b_1,...,b_s]$. This can be traced to the fact that
the singularity is resolved by a bundle which is constructed out
of $s+1$ patches with $s$ transition functions that are specified
by the numbers $b_i$. Each of these glueing steps introduces a
sphere, which in turn supports a (1,1)-form. The intersection
properties of these 2-spheres are described by Hirzebruch-Jung
trees, which for a $\mathZ/n\mathZ$ action is just an $SU(n+1)$
Dynkin diagram, while the numbers $b_i$ describe the intersection
numbers. We see from this that the resolution of a curve of genus
$g$ thus introduces $s$ additional generators to the second
cohomology group $\rmH^2(X)$, and $g\times s$ generators to the
intermediate cohomology $\rmH^3(X)$.

Hence we have shown that the cohomology of weighted hypersurfaces
is determined completely by the cohomology of curves. Because the
Jacobian, which we will describe in the next subsection, is the
only motivic invariant of a smooth projective curve this says that
for weighted hypersurfaces the main motivic structure is carried
by their embedded curves. We will come back to the motivic
structure of Calabi-Yau varieties in Section 6.

\subsection{Cohomology of weighted curves}

For smooth algebraic curves $C$ of genus $g$ the de Rham
cohomology group $\rmH^1_{\rmdR}(C)$ decomposes (over the complex
number field $\mathC$) as \beq \rmH^1_{\rmdR}(C)~\cong~\rmH^0(C,
\Om^1) \oplus \rmH^1(C,\cO).\lleq{hodge-split} The Jacobian $J(C)$
of a curve $C$ of genus $g$ can be identified with \beq
J(C)=\mathC^g/\Lambda,\eeq where $\Lambda$ is the period lattice
\beq \Lambda:= \left\{\left(\dots,\int_a \om_i,\dots
\right)_{i=1,...,g}~{\Big|}~ a \in \rmH_1(C,\mathZ),~ \om_i \in
\rmH^0(C,\Om^1) \right\},\eeq where the $\om_i$ form a basis.
Given a fixed point $p_0\in C$ on the curve there is a canonical
map from the curve to the Jacobian, called the Abel-Jacobi map
\beq \Psi: C \lra J(C), \eeq defined as \beq p \mapsto
\left(\dots, \int_{p_0}^p \om_i,\dots\right) \rmmod~ \Lambda .\eeq

We are interested in curves of Brieskorn-Pham type, i.e. curves of
the form \beq  C_d= \left\{x^d + y^a + z^b =0 \right\} \in
\IP_{(1,k,\ell)}[d],\eeq such that $a=d/k$ and $b=d/\ell$ are
positive rational integers. Without loss of generality we can
assume that $(k,\ell)=1$. The genus of these curves is given by
\beq g(C_d) = \frac{1}{2}(2-\chi) = \frac{(d-k)(d-\ell)+(k\ell
-d)}{2k\ell}. \eeq

For non-degenerate curves in the configurations
$\IP_{(1,k,\ell)}[d]$ the set of forms \beq
\rmH^1_{\rmdR}(\IP_{(1,k,\ell)}[d]) = \left\{ \om_{r,s,t}=
y^{s-1}z^{t-d/\ell}dy~{\Big |}~ r+ks+\ell t = 0~{\rmmod}~d,
~\left(\matrix{1\leq r \leq d-1, \cr ~1\leq s \leq
\frac{d}{k}-1,\cr ~1\leq t \leq \frac{d}{\ell} -1\cr}\right)
\right\}\lleq{weighted-basis} defines a basis for the de Rham
cohomology group $\rmH^1_{\rmdR}(C_d)$ whose Hodge split is given
by \bea \rmH^0\left(C_d,\Om_{\mathC}^1\right) &=&
\left\{\om_{r,s,t}~|~r+ks+\ell t=d\right\} \nn \\
\rmH^1\left(C_d,\cO_{\mathC}\right) &=&
\left\{\om_{r,s,t}~|~r+ks+\ell t=2d\right\}. \eea

In order to show this we view the weighted projective space as the
quotient of projective space with respect to the actions
$\mathZ_k:[0~1~0]$ and $\mathZ_{\ell}: [0~0~1]$, where we use the
abbreviation $\mathZ_k=\mathZ/k\mathZ$ and for any group
$\mathZ_r$ the notation $[a,b,c]$ indicates the action \beq
[a,b,c]:~(x,y,z) \mapsto (\g^ax, \g^by, \g^cz),\eeq where $\g$ is
a generator of the group. This allows us to view the weighted
curve as the quotient of a projective Fermat type curve \beq
\IP_{(1,k,\ell)}[d] = \IP_2[d]/\mathZ_k \times \mathZ_{\ell}:
\left[\matrix{0&1&0\cr 0&0&1\cr}\right].\eeq These weighted curves
are smooth and hence their cohomology is determined by considering
those forms on the projective curve $\IP_2[d]$ which are invariant
with respect to the group actions. A basis for
$\rmH^1_{\rmdR}(\IP_2[d])$ is given by the set of forms \beq
\rmH^1(\IP_2[d]) = \left\{ \om_{r,s,t}= y^{s-1}z^{t-d}dy~{\Large
|}~ 0<r,s,t<d,~~r+s+t=0~(\rmmod ~d),~~r,s,t\in \mathN
\right\}.\lleq{proj-basis}

Denote the generator of the $\mathZ_k$ action by $\a$ and consider
the induced action on $\om_{r,s,t}$ \beq \mathZ_k:~~\om_{r,s,t}
\mapsto \a^s \om_{r,s,t}.\eeq It follows that the only forms that
descend to the quotient with respect to $\mathZ_k$ are those for
which $s = 0 (\rmmod~k)$. Similarly we denote by $\beta$ the
generator of the action $\mathZ_{\ell}$ and consider the induced
action on the forms $\om_{r,s,t}$ \beq \mathZ_{\ell}:~~\om_{r,s,t}
\mapsto \beta^{t-d} \om_{r,s,t}.\eeq Again we see that the only
forms that descend to the quotient are those for which
$t=0(\rmmod~\ell)$.

\subsection{Abelian varieties from weighted Jacobians}

Jacobian varieties in general are not abelian varieties with
complex multiplication. The question we can ask, however, is
whether the Jacobians of the curves that determine the cohomology
of the Calabi-Yau varieties can be decomposed such that the
individual factors admit complex multiplication by an order of a
number field. In this section we show that this is indeed the case
and therefore we can define the complex multiplication type of a
Calabi-Yau variety in terms of the CM types induced by the
Jacobians of its curves.

It was shown by Faddeev  \cite{f61}\fnote{4}{More accessible are
the references \cite{w76} \cite{g78}, \cite{r78} on the subject.}
that the Jacobian variety $J(C_d)$ of Fermat curves $C_d\subset
\IP_2$ splits into a product of abelian factors $A_{\cO_i}$ \beq
J(C_d) \cong \prod_{\cO_i \in \cI/(\mathZ/d\mathZ)^{\times}}
A_{\cO_i}, \eeq where the set $\cI$ provides a parametrization of
the cohomology of $C_d$, and the sets $\cO_i$ are orbits in $\cI$
of the multiplicative subgroup $(\mathZ/d\mathZ)^{\times}$ of the
group $\mathZ/d\mathZ$. More precisely it was shown that there is
an isogeny \beq i: J(C_d) \lra \prod_{\cO_i \in
\cI/(\mathZ/d\mathZ)^{\times}} A_{\cO_i},\eeq where an isogeny $i:
A \ra B$ between abelian varieties is defined to be a surjective
homomorphism with finite kernel. In the parametrization used in
the previous subsection $\cI$ is the set of triplets $(r,s,t)$ in
(\ref{proj-basis}) and the periods of the Fermat curve have been
computed by Rohrlich \cite{r78} to be \beq \int_{A^jB^k\kappa}
\om_{r,s,t} = \frac{1}{d} B\left(\frac{s}{d},\frac{t}{d}\right)
(1-\xi^s)(1-\xi^t)\xi^{js+kt}, \eeq where $\xi$ is a primitive
$d-$th root of unity, and \beq B(u,v) = \int_0^1
t^{u-1}(1-v)^{v-1}dt \eeq is the classical beta function. $A,B$
are the two automorphism generators \bea A(1,y,z) &=& (1, \xi y, z) \nn \\
B(1,y,z) &=& (1, y,\xi z)\eea and $\kappa$ is the generator of
$\rmH_1(C_d)$ as a cyclic module over $\mathZ[A,B]$. The period
lattice of the Fermat curve therefore is the span of \beq
\left(\dots, \xi^{jr+ks}(1-\xi^r)(1-\xi^s) \frac{1}{d}
B\left(\frac{r}{d},\frac{s}{d}\right), \dots
\right)_{\stackrel{1\leq r,s,t \leq d-1}{r+s+t=d}},~~ \forall
0\leq j,k\leq d-1. \eeq

The abelian factor $A_{[(r,s,t)]}$ associated to the orbit
$\cO_{r,s,t}=[(r,s,t)]$ can be obtained as the quotient \beq
A_{[(r,s,t)]} = \mathC^{\vphi(d_0)/2}/\Lambda_{r,s,t}, \eeq where
$d_0 = d/\rmgcd(r,s,t)$ and the lattice $\Lambda_{r,s,t}$ is
generated by elements of the form \beq
\si_a(z)(1-\xi^{as})(1-\xi^{at})
\frac{1}{d}B\left(\frac{<as>}{d},\frac{<at>}{d}\right), \eeq where
$z\in \mathZ[\mu_{d_0}]$, $\si_a \in
\rmGal(\mathQ(\mu_{d_0}/\mathQ)$ runs through subgroups of the
Galois group of the cyclotomic field $\mathQ(\mu_{d_0})$ and $<x>$
is the smallest integer $0\leq x <1$ congruent to $x$ mod $d$.

We adapt this discussion to the weighted case. Denote the index
set of triples $(r,s,t)$ parametrizing the one-forms of the
weighted curves $C_d \in \IP_{(1,k,\ell}[d]$ again by $\cI$. The
cyclic group $(\mathZ/d\mathZ)^{\times}$ again acts on $\cI$ and
produces a set of orbits \beq \cO_{r,s,t} =  [(r,s,t)] \in
\cI/(\mathZ/d\mathZ)^{\times}.\eeq Each of these orbits leads to
an abelian variety $A_{[(r,s,t)]}$ of dimension \beq \rmdim
A_{[(r,s,t)]} = \frac{1}{2} \vphi\left(d_0\right) , \eeq and
complex multiplication with respect to the field $F_{[(r,s,t)]} =
\mathQ(\mu_{d_0})$, where $d_0=d/\rmgcd(r,ks,\ell t)$. This leads
to an isogeny \beq i: J(\IP_{(1,k,\ell)}[d]) ~\lra
~\prod_{[(r,s,t)]\in \cI/(\mathZ/d\mathZ)^{\times}}
A_{[(r,s,t)]}.\eeq

 The complex multiplication type of the abelian factors
 $A_{r,s,t}$ of the
Jacobian $J(C)$ can be identified with the set \beq \rmH_{r,s,t}
:= \left\{a\in (\mathZ/d\mathZ)^{\times}~|~<ar>+<aks>+<a\ell
t>=d\right\} \eeq via a homomorphism from $\rmH_{r,s,t}$ to the
Galois group. More precisely, the CM type is determined by the
subgroup $G_{r,s,t}$ of the Galois group of the cyclotomic field
that is parametrized by $\rmH_{r,s,t}$ \beq
 G_{r,s,t} = \left\{\si_a \in \rmGal(\mathQ(\mu_{d_0})/\mathQ)~|~a
 \in \rmH_{r,s,t}\right\} \eeq
by considering \beq (F,\{\phi_a\}) = (\mathQ(\mu_{d_0}), \{\si_a
~|~ \si_a \in G_{r,s,t}\}).\eeq

\vskip .4truein

\section{Arithmetic of Abelian Varieties}

Abelian varieties with complex multiplication have special
properties because of their particular symmetries. It turns out
that even though the theory of CM fields associated to higher
dimensional varieties is not as complete as the theory associated
to elliptic curves with complex multiplication, a number of key
results of the elliptic theory have been generalized to abelian
varieties, mostly by Shimura.

Suppose that the abelian variety $A$ of dimension $n$ has complex
multiplication by the ring of integers $\cO_F$ of some CM field
$F$ (or by an order in $F$), and that the CM type of the variety
is given by $(F,\{\phi_i\}_{i=1,...,n})$. The arithmetic structure
induced by higher dimensional varieties is concerned not with $F$
itself but the so-called reflex field $\hatF$, which depends not
only on the field $F$, but also on the CM type of $F$. $\hatF$ is
defined as the extension $\mathQ(\sum_{i=1}^n a^{\phi_i})$ of the
field of rational numbers by adjoining certain traces of elements
$a\in F$. The higher dimensional analog of the elliptic field of
moduli then gives an unramified abelian extension of the field
$\hatF$ \cite{s71}. Even though ramified class fields over $\hatF$
can be obtained as well, the theory leads to less complete results
because it does not give all abelian extensions of $\hatF$. For
this reason Hilbert's twelfth problem is still not solved for CM
fields.

The basic question is whether there is some simple way to
characterize the kind of subfield of the maximal abelian extension
$F_{\rmab}$ of a CM field $F$ that can be obtained by adjoining to
the reflex field the moduli fields of abelian varieties as well as
the points of finite order. A nice result in this direction has
been obtained by Wei \cite{w94}. In brief, her theorem states that
given a CM field $F$ with totally real subfield $F_{\IR}$, the
field $F_{\rmmod}$ generated by the moduli and torsion points of
all polarized abelian varieties of CM type whose reflex field is
contained in $F$, is the subfield of $F_{\rmab}$ that is fixed
under the subgroup $H$ of the Galois group $\rmGal(F_{\rmab}/F)$
generated by the verlagerungs map \beq \rmVer:
\rmGal(\bmathQ/F_{\IR})_{\rmab} ~\lra ~\rmGal(\bmathQ/F).\eeq More
concisely,  \beq F_{\rmmod} = (F_{\rmab})^H.\eeq The verlagerungs
map $\rmVer$ involved here is a general construction which assigns
to a subgroup $H$ of a group $G$ a homomorphism between the
abelianizations $G_{\rmab}=G/(G,G)$ and $H_{\rmab}=H/(H,H)$ of the
pair of groups \beq \rmVer: G_{\rmab} ~\lra ~H_{\rmab}.\eeq
Consider a system of $C$ of representatives for the left cosets of
$H$ in $G$. For each $g\in G$ decompose the translate $ga$ for any
$a\in C$ as $ga=a'g_a$, with $g_a\in H$ and $a'\in C$. The
verlagerungs map is then defined as \beq \rmVer(g~\rmmod~(G,G)) =
\prod_{a\in C} g_a~\rmmod~(H,H).\eeq More details can be found in
\cite{n99,rv98}.

We see from this that, even though the results are weaker, the
generalization from the imaginary quadratic fields of
one-dimensional abelian varieties to the CM fields of higher
dimensions allows for a fairly nice characterization.

\vskip .4truein

\section{Deligne's Period Conjecture}

Our focus in this paper is on the fields of complex multiplication
that we derive from the abelian varieties which we construct from
Calabi-Yau varieties. In Moore's analysis of higher dimensional
manifolds the focus is on fields derived from the periods of the
variety. In this section we describe how the period approach can
be recovered from our higher-dimensional complex multiplication
point of view via Deligne's conjecture formulated in \cite{d79}.
Precursors to Deligne's formulation can be found in Shimura's work
\cite{s75,s76,s77}.

Deligne's conjecture in its motivic formulation is also useful in
the present context because it allows us to provide a general
perspective for our results which will furnish what we expect to
be a useful general framework in which to explore further the
arithmetic nature of attractor varieties. Motives are somewhat
complicated objects whose status is reminiscent of string theory:
different realizations are used to probe what is believed to be
some yet unknown unifying universal cohomology theory of varieties
which satisfies a number of expected functorial properties. More
precisely, motives are characterized by a triplet of different
cohomology theories together with a pair of compatibility
homomorphisms. In terms of these ingredients a motive then can be
described by the quintuplet of objects \beq (\rmM_{\rmB},
\rmM_{\rmdR}, \rmM_{\ell}, \rmI_{\rmB, \si}, \rmI_{\ell,
\bsi}),\eeq where the three first entries are cohomological
objects constructed via Tate twists from the Betti, de Rham, and
\'etale cohomology, respectively. Furthermore $\rmI_{\rmB,\si}$
describes a map between the Betti and de Rham cohomology, while
$\rmI_{\ell, \bsi}$ is a map between Betti and \'etale
cohomology\fnote{2}{Detailed reviews of motives can be found in
\cite{jks94}.}. The focus in the present paper is mostly on
motives derived from the first (co)homology groups $\rmH^1(A)$ and
$\rmH_1(A)$ of abelian varieties $A$, as well as the primitive
cohomology of Fermat hypersurfaces.

The second ingredient in Deligne's conjecture is the concept of an
$L-$function. This can be described in a number of equivalent
ways. Conceptually, the perhaps simplest approach results when it
can be derived via Artin's zeta function as the Hasse-Weil
L-function induced by the underlying variety, i.e. by counting
solutions of the variety over finite
fields\fnote{3}{Ref.\cite{s95} contains a brief description of
this construction.}. The complete L-function receives
contributions from two fundamentally different factors,
$\Lambda(\rmM,s) = L_{\infty}(\rmM,s) L(\rmM,s)$. The infinity
term $L_{\infty}(\rmM,s)$ originates from those fields over which
the underlying variety has bad reduction, i.e. it is singular,
while the second term $L(\rmM,s)$ collects all the information
obtained from the finite fields over which the variety is smooth.
The complete L-function is in general expected to satisfy a
functional equation, relating its values at $s$ and $1-s$. A
motive is called critical if neither of the infinity factors in
the functional equation has a pole at $s=0$.

The final ingredient is the concept of the period of a motive, a
generalization  of ordinary periods of varieties. Viewing the
motive $\rmM$ as a generalized cohomology theory, Deligne
formulates the notion of a period $c^+(\rmM) \in
\mathC^{\times}/\mathQ^{\times}$ by taking the determinant of the
compatibility homomorphism \beq I_{\rmB,\si}: \rmM_{\rmB} \lra
\rmM_{\rmdR} \eeq between the Betti and the deRham realizations of
the motive $\rmM$. Deligne's basic conjecture then relates the
period and the L-function  via $L(\rmM,0)/c^+(\rmM) \in \mathQ$.
Contact with the Hasse-Weil L-function is made by noting that for
motives of the type $\rmM=\rmH(X)(m)$ with Tate twists one has
$L(\rmM,0)=L(X,m)$.

Important for us is a generalization of this conjecture which
involves motives with coefficients. Such motives can best be
described via algebraic Hecke characters, which are of particular
interest for us because they come up in the L-function of
projective Fermat varieties. Algebraic Hecke characters were first
introduced by Weil and called Hecke characters of type $A_0$,
which is what they are called in the older literature. The beauty
of algebraic Hecke characters is that one is immediately led to a
clear distinction between the defining field $K$ and the field in
which a character lives, i.e. the field $F$ of values. In the
context of motives constructed from these characters the field $F$
becomes the field of complex multiplication. Deligne's conjecture
emerges in the following way.

{\bf Deligne Period Conjecture:} \beq \frac{L(\rmM,0)}{c^+(\rmM)}
\in F.\eeq

This shows why the Deligne conjecture is of interest to us. The
period and the L-function determine the same field, which is the
CM field of the motive. Deligne's conjecture has been proven by
Blasius for Fermat hypersurfaces \cite{b86}.

\vskip .4truein

\section{Summary and Generalizations}

We have shown that the concepts used to describe attractor
varieties in the context of elliptic compactifications can be
generalized to Calabi-Yau varieties with finite fundamental
groups. We have mentioned above that the abelian property is
neither carried by the variety itself nor the generalized
intermediate Jacobian \beq J^n(X) =
\rmH^{2n-1}(X_{\rman},\mathC){\Large
/}\rmH^{2n-1}(X_{\rman},\mathZ(n)) +
\rmF^n\rmH^{2n-1}(X_{\rman},\mathC),\eeq  but by the Jacobians of
the curves that are the building blocks of the middle-dimensional
cohomology $\rmH^{\rmdim_{\mathC} X}(X)$. These Jacobians
themselves do not admit complex multiplication, unlike the
situation in the elliptic case, but instead split into different
factors which admit different types of complex multiplication, in
general. Furthermore the ring class field can be generalized to be
the field of moduli, and we can consider also points on the
abelian variety that are of finite order, i.e. torsion points, and
the field extensions they generate.

This allows us to answer a question posed in \cite{m98} which
asked whether the absolute Galois group $\rmGal(\bK/K)$ could play
a role in the context of $N=2$ compactifications of type IIB
strings. This is indeed the case. Suppose we have given an abelian
variety $A$ defined over a field $K$ with complex multiplication
by a field $F$. Then there is an action of the absolute Galois
group $\rmGal(\bK/K)$ of the closure $\bK$ of $K$ on the torsion
points of $A$. This action is described by a Hecke character which
is associated to the fields $(K,F)$ \cite{s71}.

We have mentioned already that in general the (Griffiths)
intermediate Jacobian is only a torus, not an abelian variety.
Even in those cases it is however possible to envision the
existence of motives via abelian varieties associated to a variety
$X$. Consider the Chow groups $\rmCH^p(X)$ of codimension $p$
cycles modulo rational equivalence and denote by
$\rmCH^p(X)_{\rmhom}$ the subgroup of cycles homologically
equivalent to zero. Then there is a homomorphism, the Abel-Jacobi
homomorphism, which embeds $\rmCH^p(X)_{\rmhom}$ into the
intermediate Jacobian \beq \Psi: \rmCH^p(X)_{\rmhom} \lra
J^p(X).\eeq The image of $\Psi$ on the subgroup $\cA^p(X)$ defined
by cycles algebraically equivalent to zero does in fact define an
abelian variety, even if $J^p(X)$ is not an abelian variety but
only a torus \cite{s79}. Hence we can ask whether attractor
varieties are distinguished by Abel-Jacobi images which admit
complex multiplication.

Even more general, we can formulate this question in the framework
of motives because of Deligne's conjecture. Thinking of motives as
universal cohomology theories, it is conceivable that attractor
varieties lead to motives in the abelian category with (potential)
complex multiplication. The standard cycle class map construction
of $\rmCH^p(X)_{\rmhom}$ is replaced by the first term of a
(conjectured) filtration in the resulting K-theory.

Combining the two threads of our analysis illustrates that the two
separate discussions in \cite{m98} characterizing toroidal
attractor varieties via complex multiplication on the one hand,
and Calabi-Yau hypersurfaces via periods on the other, are two
aspects of our way of looking at this problem. This is the case
precisely because of Deligne's period conjecture which relates the
field of the periods to the field of complex multiplication via
the L-function of the variety (or motive). Thus a very pretty
unified picture emerges.

\vskip .5truein

{\bf {\large Acknowledgement}}

We are grateful to G. Moore and P. Deligne for discussions. ML and
RS thank the Kavli Institute for Theoretical Physics, Santa
Barbara, for hospitality and support through KITP Scholarships
during the course of part of this work. RS would also like to
thank the organizers of the Duality Workshop at the ITP in 2001.
This research was supported in part by the National Science
Foundation under grant No. PHY99-07949. The work of M.L. and R.S.
has been supported in part by NATO under grant CRG 9710045. The
work of VP was supported by NSF under grant No. PHY98-02484.

\end{document}